\documentclass[aps,prd,showpacs,preprintnumbers]{revtex4}
\usepackage{graphics}
\def \beq{\begin{equation}}
\def \eeq{\end{equation}}
\def \beqa{\begin{eqnarray}}
\def \eeqa{\end{eqnarray}}
\def \lqcd{\Lambda_{QCD}}

\def \etal{{\sl et al.\/}}
\def \jhep{{\sl J.\ H.\ E.\ P.\/}}
\def \np{{\sl Nucl.\ Phys.\/}}
\def \pl{{\sl Phys.\ Lett.\/}}
\def \pr{{\sl Phys.\ Rev.\/}}
\def \prl{{\sl Phys.\ Rev.\ Lett.\/}}
\def \eg{{\sl e.\ g.\/}}
\def \ie{{\sl i.\ e.\/}}
\def \etc{{\sl etc.\/}}

\begin{document}
 
\title{The phase diagram of QCD at small chemical potentials}
\author{Sourendu \surname{Gupta}}
\email{sgupta@tifr.res.in}
\affiliation{Department of Theoretical Physics, Tata Institute of Fundamental
         Research,\\ Homi Bhabha Road, Mumbai 400005, India.}

\begin{abstract}
We investigate the phase diagram of QCD at small chemical potentials,
\ie, when chiral and flavour symmetry breaking involves the pairing of
a quark and an antiquark. The phase diagram of two-flavour QCD at small
chemical potentials involves chiral symmetry restoration and charged
pion condensation. We extend previous studies
of the topology of the phase diagram, in sections with high degree
of symmetry, to the physical case of fully broken flavour symmetry,
using generic thermodynamic arguments. We argue that the extension is
unique and present the result. In three flavour QCD the
phase diagram for chiral symmetry restoration is less well constrained.
However, we argue that present lattice data allows just two
different phase diagrams, which we discuss.
\end{abstract}
\preprint{TIFR/TH/07-0?}
\maketitle

\section{Introduction}

More than thirty years after the first discussions about a phase
transition in QCD \cite{transition}, only small portions of the phase
diagram have been explored. Although the complete phase diagram of
QCD is of high dimensionality, experiments can at best explore a three
dimensional section of the full phase diagram. Further, collisions of
heavy-ions have only a single control parameter, the CM energy, $\sqrt
S$. As a result they explore a single line in this three-dimensional
phase diagram. By varying the nuclei being collided, one could, perhaps,
extend the search to a small patch around the line.  The field is wide
open for new ideas on experimental coverage of the QCD phase diagram.

Theoretical work is no less constrained by the tools of the trade. In
regions of high symmetry (for example, when the quark masses vanish),
universality arguments \cite{pw} have been used to put constraints on
the phase diagram. Such arguments are realized in models, for example,
effective meson models, four-Fermi models or random matrix models, which
have the same symmetries as QCD. The resulting predictions of universal
properties, \eg, the order of the transition and critical indices, are
expected to coincide with QCD. Since the locations of phase transitions
are not universal, models should be used to constrain the topology of
the phase diagram rather than quantitative predictions of the location of
phase boundaries or critical points.  When the symmetries are broken, as
in the real world, the usefulness of these models is curtailed further.
Weak coupling methods for QCD give precise quantitative predictions,
but for high temperatures and densities, when the QCD coupling is small
enough.  Lattice computations were long confined to the region with
vanishing chemical potentials, extensions to finite chemical potential
being constrained by the fermion-sign problem. The first systematic
non-perturbative treatments of QCD at non-vanishing chemical potential
using lattice methods have now begun, and the first results are now
available \cite{lattice}. In spite of these limitations, tremendous
progress has been made. As we demonstrate in this paper, known results
can now be uniquely exended, using only thermodynamic arguments, to yield
the topology of the full phase diagram of two flavour QCD and strongly
constrain it for three flavours.

The topology of the phase diagram of QCD is constrained by its
symmetries.  It is well-known that QCD possesses a set of approximate
global symmetries, called flavour symmetries, related to the phases
of quark wavefunctions. QCD with two flavours of massless quarks
would possesses a chiral symmetry $SU_L(2)\times SU_R(2)$ ($L$ and $R$
are transformations on left and right handed quarks respectively).
The up and down quark masses ($m_u$, $m_d$ respectively) break this
symmetry. Since $m=(m_u+m_d)/2$ is non-zero the chiral symmetry is broken
to the diagonal vector symmetry, $SU_V(2)$, called isospin.  Since $m$
is of the order of a few MeV, and much smaller than the scale, $\lqcd$,
chiral symmetry is approximately valid, being broken at the level of
5--10\%. The mass difference $\Delta m=m_d-m_u$ is non-zero but small,
thus violating isospin symmetry by a small amount.  There is also a
semi-light quark flavour, the strange, which has mass, $m_s$, comparable
to $\lqcd$. Including this extends the chiral group to $SU_L(3)\times
SU_R(3)$ \cite{dashen}.  According to data, this symmetry is broken at
the 25\% level down to the two-flavour symmetry $SU_L(2)\times SU_R(2)$
\cite{broken}.  In addition to these approximate flavour symmetries
there is also the exact phase symmetry $U_B(1)$ whose charge is the
baryon number. The axial phase symmetry, $U_A(1)$, is broken at $T=0$ by
instantons. There is mounting evidence that this symmetry is not restored
through a phase transition \cite{axial}. In this paper we shall assume that
there is no $U_A(1)$ restoring phase transition in QCD.

The large global symmetries of QCD can be broken in many ways, thus giving
rise to a complicated phase diagram. In this paper we confine ourselves
to the phase diagram at small chemical potentials. By this we mean that
the order parameters involve pairing of quarks and antiquarks. At larger
chemical potentials there are other interesting phases where the pairing
could be between two quarks \cite{rw}. We do not examine these phases
in this work. A range of intermediate chemical potentials may exist where
both quark-antiquark and quark-quark pairings need to be taken into
account \cite{baym}. If this is so, then some details of the phase diagrams
presented here would have to be extended.

Since the phase diagram is structured around the breaking of chiral and
flavour symmetry, one might expect that it is independent of the number of
colours, $N_c$. This is correct for $N_c\ge3$.  For the specific case of
$N_c=2$, however, the fact that quark representations are real means that
the chiral symmetry is enhanced. As a result, the considerations below
do not apply to $N_c=2$.  There is a large body of literature on this
particular case, and we refer the interested reader to a recent review
\cite{nctwo}. Interestingly enough, as $N_c\to\infty$ the fact that a
baryon contains $N_c$ quarks implies that the region of small chemical
potentials, in the technical sense adopted here, increases to $\mu\propto
N_c$. As a result, several interesting new phases open up in the hadronic
regime and may be studied using different order parameters \cite{largen}.

We argue here that thermodynamic considerations allow us to extend
presently available knowledge to large parts of the parameter space of
QCD and enable us to build a qualitative picture of the complete phase
diagram of QCD for small chemical potential. We deduce the topology of
the three dimensional slice of the phase diagram of two-flavour QCD
which may be accessible to experimental tests. We also indicate how
these arguments allow us to constrain the phase diagram of QCD with up,
down and strange quarks.

The plan of this paper is the following: in the next section we briefly
review the Gibbs' phase rule in the form that we will use it. The two
sections following that deal with $N_f=2$ and 3 respectively. The final
section contains a summary of our results.

\section{The Gibbs' phase rule}\label{sc.gibbs}

In this paper we investigate the phase diagram of QCD using an essential
tool of thermodynamics: the Gibbs' phase rule. Before stating the rule
we recall that a system in thermodynamic equilibrium is fully described
by a certain number of extensive thermodynamic quantities. This set of
extensive quantities, among which we must always count the entropy $S$
and the energy $E$, serve as coordinates in the so-called Gibbs space.
The thermodynamically stable states of a system are in one-to-one
correspondence to a convex surface $E(S,\cdots)$ in Gibbs space.
The thermodynamic intensive quantities are derivatives of $E$ with respect
to one of the other extensive quantities, the derivative with respect to
$S$ being $T$. By the process of taking Legendre transforms of $E$ with
respect to each of the intensive variables, one reaches a description of
thermodynamics in terms of the intensive quantities and a free energy,
$\cal G$, which is extensive.  A phase diagram is obtained by projecting
out information on $\cal G$, and describes the regions in which different
phases of a system are thermodynamically stable.

By construction almost all points in the phase diagram correspond to
one pure phase. This is the one with the lowest free energy among all possible
phases. If the system has more than one phase, then at some points in
the phase diagram two phases may coexist. If we label the phases by $a$
and $b$, then the condition for coexistence is the equality of free energy
densities in the two phases,
\beq
   g_a(T,\mu_u,m_u,\cdots) = g_b(T,\mu_u,m_u,\cdots),
\label{coexist}\eeq
If the dimension of the phase diagram is $D$, then two phases
coexist along solutions of the above equation, \ie, generically along
hypersurfaces of $D-1$ dimensions.  Three-phase coexistence requires
simultaneous equality of three free energies, and hence occurs along
hypersurfaces of $D-2$ dimensions. This surface of 3-phase coexistence
is clearly the intersection of three surfaces of 2-phase coexistence,
obtained by taking the phases pairwise.  $\cal P$ phases generically
coexist along hypersurfaces of $D+1-\cal P$ dimensions, which are the
intersection of $\cal P$ surfaces of two-phase coexistence.  This is
one form of the statement of Gibbs' phase rule.

A two-phase coexistence surface either has no boundary or ends
in a surface of one lower dimension, $D-2$, called a critical
surface. Similarly, a 3-phase coexistence surface may have a
boundary. Such a boundary is a $D-3$ dimensional surface called a
tricritical surface. Since the 3-phase coexistence surface is the
intersection of three 2-phase coexistence surfaces, the tricritical
surface is the intersection of three critical surfaces \cite{griffiths}.
Boundaries of $\cal P$-phase coexistence surfaces, when they exist, are
called $\cal P$-critical surfaces, and can be viewed as intersections of
$\cal P$-critical surfaces.  Clearly a $\cal P$-critical
surface is one at which $\cal P$ different phases simultaneously become
indistinguishable. Thus another form of Gibbs' phase rule states that
when the dimension of the phase diagram is $D$, then there may be
$D-1$ dimensional surfaces of two-phase coexistence (first order phase
transitions), $D-2$ dimensional critical surfaces, $D-3$ dimensional
tricritical surfaces, \etc.

The most well-known application of the rule is to phase diagrams of
chemically pure substances (for example, water), characterised by two
intensive parameters $T$ and the pressure $P$. Since $D=2$, one has lines
of two-phase coexistence. Such lines can either end in critical points,
or two such lines can meet at isolated triple points (\ie, points of
three phase coexistence). Both these possibilities are realized in the
well-known phase diagram of water. A different example is in mixtures of
two chemically pure substances (water-alcohol, binary alloys, QCD with
quarks, \etc), where an additional intensive parameter, the chemical
potential $\mu$, makes the phase diagram three dimensional. In this case one
has surfaces of two-phase coexistence, bounded by critical lines. Three
surfaces can intersect along lines of three-phase coexistence. The end
point of such a line is a tricritical point, and one can view this point
also as the crossing point of three critical lines. The realization of
these possibilities in ${\rm He}^3$-${\rm He}^4$ mixtures was treated
in \cite{griffiths}.

If the phase diagram is known only in some part of the space of intensive
variables where calculations are tractable, then the Gibbs' phase rule
allows us to constrain the possibilities that arise from extrapolations to
larger parts of the space. As we show in the coming sections, the constraints
are enormous, and sometimes they determine the topology of the phase diagram
completely.

In QCD a choice of the extensive quantities can be the order parameters
for the breaking of chiral symmetry (one condensate for each flavour
of quarks and the net number of quarks of each flavour). Conjugate
to these are the couplings--- the quark masses and the
chemical potentials respectively. Adding to this set
the total energy, one sees that for $N_f=2$ the phase diagram has 5
dimensions. For $N_f=3$ the strange quark mass and chemical potential
are added, as a result of which the phase diagram is 7 dimensional.
Experiments are constrained to work with given quark masses, and hence
explore a phase diagram of dimension 3 for $N_f=2$ and dimension 4 for
three flavours.

This is true if one examines a strongly interacting system on a time
scale much larger than the slowest strong interaction related relaxation
time, but much smaller than the time scale of the flavour changing
weak interactions. This is the case in heavy-ion collisions, where the
expanding fireball cools and freezes out on time scales of the order of a
few fermis, whereas the typical strangeness changing time scale is many
orders of magnitude larger. This might seem to indicate that the full
4 dimensional phase diagram can be explored. However, in the initial
state the only conserved quantum numbers are the baryon number, $B$,
and the electrical charge, $Q$. Since these two, and the total energy,
are the only quantities which can be manipulated in experiments, the three
dimensional slice of the phase diagram conjugate to these variables is the
only part of the full phase diagram which is accessible to any possible
experiment. As is well-known, other parts of the $N_f=3$ phase diagram
may become accessible in situations at high densities and relatively
low temperatures, where Fermi-blocking of the light quark states may
increase the lifetime of strange quarks. This situation may well be
realized in compact stellar objects. In the early universe, around the
time of the QCD phase transition, the baryon density is much higher than
now, but still small enough that $\mu_B\ll T$. Also, the inverse Hubble
time is much smaller than it is now, but much larger than weak the weak
interaction time scale. Hence, in the early universe, both strong and
weak interactions should be considered to be in equilibrium, so that
the phase diagram reverts to being 3 dimensional. Unfortunately, the
chemical potentials in the early universe are so small that essentially
only the temperature axis is explored.

\section{Two flavours}

For $N_f=2$ QCD the five intensive quantities can be chosen to be $T$,
$\mu_u$, $\mu_d$, $m_u$ and $m_d$. A common alternative choice is to
use $\mu_B$, conjugate to the baryon number, $B$, and $\mu_I$, conjugate
to the third component of isospin, $I_3$. Moving between the ensembles
corresponding to these is straightforward; detailed formulae are given
in \cite{quasiquarks}. It is useful to note that these intensive quantities
are conjugate to the extensive thermodynamic variables
\beq
   N_u = \frac{\partial{\cal G}}{\partial\mu_u},
   \qquad {\mathbf s}_u = \frac{\partial{\cal G}}{\partial m_u},
\label{extens}\eeq
and similar relations for down quarks, where $N_u$ is the net number of
up quarks and ${\mathbf s}_u$ is an extensive quantity whose density is
the up quark condensate $\langle\overline uu\rangle$.

\subsection{Finite temperature}

The best-known part of the phase diagram for $N_f=2$ QCD is the
temperature axis for $m=\Delta m=\mu_B=\mu_I=0$ \cite{pw}.  The action has
the chiral symmetry $SU_L(2)\times SU_R(2)$. 
At $T=0$ this is spontaneously broken to the isospin part,
$SU_V(2)$, resulting in three massless pions arising as the
Goldstone bosons of the broken symmetry.  In the high temperature limit
the symmetry is restored. The symmetry of the theory is isomorphic to
O(4), which is known to have a second order phase transition. Universality
then indicates that there is a critical temperature, $T_c$, at which
a second order phase transition occurs from the chiral symmetry broken
phase to the symmetric phase with critical exponents in the O(4) symmetry
class. The order parameter is the isoscalar chiral condensate, ${\mathbf
s}=({\mathbf s}_u+{\mathbf s}_d)/2$, which changes from a non-zero value
at low temperatures to zero at $T_c$. For $T<T_c$ the
order parameter is negative for any positive $m$, and flips sign when $m$
in negative. Hence there is a line of first order phase transitions for
$m=0$ and $T<T_c$, which ends in the critical point at $T_c$.  There is
no finite temperature phase transition at finite $m$.

Lattice computations are performed at finite quark mass, and are consistent
with this picture since they observe only a cross over \footnote{The recent
claim that a first order transition exists \cite{pica} requires further
substantiation}. Chiral
extrapolations have given mixed results: there is no agreement on
the critical exponents which are seen. This could be a finite lattice
spacing artifact. Several studies have used staggered quarks which,
for $m=0$, have only an $U(1)\times U(1)$ chiral symmetry at finite
lattice spacing. The chiral transition in this case should lie in the
$O(2)$ universality class. It turns out to be hard to distinguish O(4)
and O(2) exponents \cite{engels}. Lattice studies are consistent with
both \cite{stagce}.  Studies with Wilson quarks show results consistent
with critical exponents in the $O(4)$ universality class \cite{cppacs}.
The situation can be clinched in simulations with dynamical overlap
quarks, which realize the full chiral symmetry. However, these simulations
are currently in their infancy.

\begin{figure}[!tbh]
   \begin{center}
   \scalebox{0.7}{\includegraphics{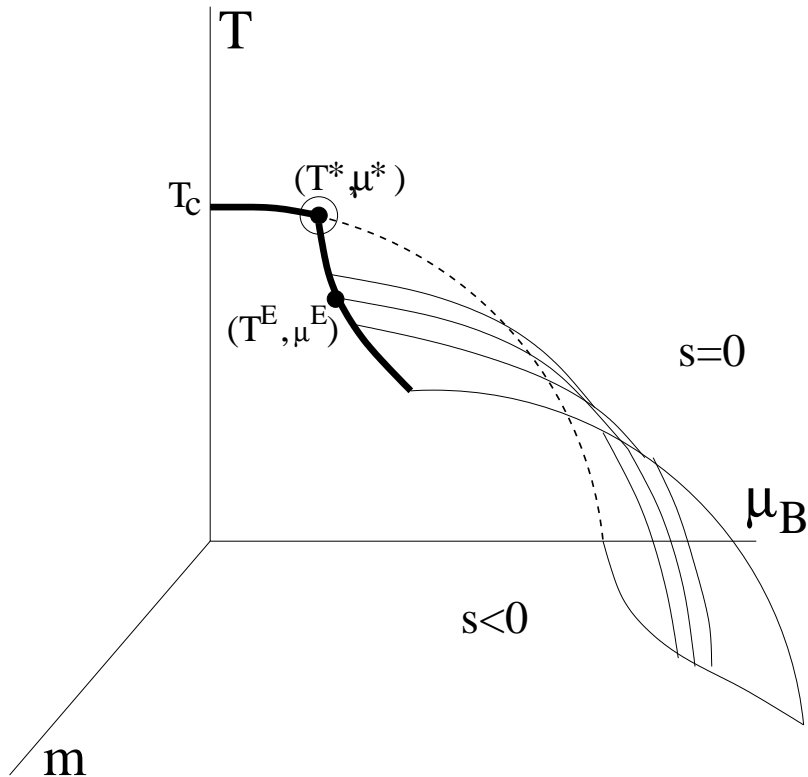}}
   \end{center}
   \caption{The phase diagram in the $T$-$\mu_B$-$m$ space \cite{kr} (only
      the side for $m\ge0$ is shown; a symmetric part of the figure for
      $m\le0$ is obtained by switching the sign of the order parameter
      $\mathbf s$). A critical line in the $O(4)$ universality class
      starts from the chiral phase transition, $T_c$ and is the boundary
      of the first order surface separating the phases with opposite signs
      of $\mathbf s$. The tricritical point at $(T^*,\mu^*)$ for $m=0$
      lies at the end of the triple line (dashed curve). At the physical
      quark mass one expects a critical end point, $(T^E,\mu^E)$, in the
      Ising universality class. This lies on the critical line which is
      the boundary of the first order surface separating the phases with
      non-vanishing and zero value of the order parameter.}
\label{fg.phsdgt0m}\end{figure}

\subsection{The $T$-$\mu_B$-$m$ section with and without isospin symmetry}

Recently attention has been focused on the three dimensional section of
the phase diagram with varying $T$, $\mu_B$ and $m$ and fixed $\Delta
m=\mu_I=0$. Using the order parameter $\mathbf s$ one finds the phase
diagram shown in Figure \ref{fg.phsdgt0m}. The plane $\mu_B=0$ was
described in the previous subsection. The first order line and its O(4)
critical end point described there extends naturally to three dimensions,
becoming a surface of 2-phase coexistence (corresponding to a sign flip
in $\mathbf s$) with a critical line as boundary in the $m=0$ plane. For
positive $m$ there is a 2-phase coexistence surface across which a large
negative value of $\mathbf s$ changes discontinuously to a smaller
value. For $m<0$ there is a symmetrically placed coexistence surface
with the sign of $\mathbf s$ reversed. These three 2-phase coexistence
surfaces intersect in a triple line in the $m=0$ plane. Since there is
a critical line bounding one of these surfaces, it must meet the triple
line at a tricritical point. We are led to the conclusion that all three
2-phase coexistence surfaces are bounded by critical lines, and the
tricritical point, at $(T^*,\mu_B^*)$, is their common intersection point.

One fact about this phase diagram is worth emphasizing. The
quantity $\mathbf s$ is an order parameter related to symmetry only
on the $m=0$ plane in Figure \ref{fg.phsdgt0m}, since it is exactly
zero everywhere above the triple-phase and critical lines. At finite
$m$, above the coexistence surface it is not exactly zero, but has a
value proportional to $m$.  Below this surface it has a much larger
value. Across the coexistence surface it has a discontinuity, and hence
can be used to flag a first order phase transition even at finite
$m$. However, along the wing critical lines, $\mathbf s$ is not an
order parameter in the sense of being zero in one phase and non-zero
in another. This is related to the fact that, for $m\ne0$, the phase
transition is not related to a symmetry.  However, along the critical
line there are long range correlations between local fluctuations in
$\mathbf s$, or, equivalently, through a fluctuation-dissipation theorem,
there are divergent susceptibilities of $\mathbf s$. This kind of critical
behaviour is said to be a liquid-gas transition, and 
expected to be in the Ising universality class.

The slopes of the coexistence surface can be determined by a
generalization of the Clapeyron-Clausius equations. For the case at hand
we can write
\beq
   d{\cal G}_i = S_i dT + B_i d\mu_B + {\mathbf s}_i dm,
\label{cceq}\eeq
where the subscript $i$ refers to the two phases (say, a and b) which
are at equilibrium along the coexistence surface. As one moves along
the surface, the changes $dT$, $d\mu_B$ and $dm$ are related by the
fact that the free energy density is the same in the two phases, eq.\
(\ref{coexist}).  In order to define a density in a relativistic theory,
one notes that all four extensive quantities scale similarly, and hence
any one of them can be used as a normalization. In determining the slope
of the coexistence surface, one has to hold one of the three intensive
quantities fixed, so one can choose to use the extensive quantity
conjugate to it to perform the normalization.

Along lines of constant $m$, one can then normalize all extensive quantities
by $\mathbf s$ to get the relation
\beq
   \left.\frac{d\mu_B}{dT}\right|_m =
     -\frac1B\,\frac{S_a/{\mathbf s}_a-S_b/{\mathbf s}_b}{
       1/{\mathbf s}_a-1/{\mathbf s}_b}.
\label{mutslope}\eeq
Since the right hand side is independent of the sign of the chiral
condensate, one can use the magnitude $|{\mathbf s}_{a,b}|$ in the
expression. Choose a to be the low temperature phase and b to be the high
temperature phase. Then, as is shown by lattice computations, $S_a<S_b$
and $|{\mathbf s}_a|>|{\mathbf s}_b|$.  As a result, the expression on
the right is negative, implying that with increasing $T$ the surface of
coexistence moves to smaller $\mu$, as shown in Figure \ref{fg.phsdgt0m}.

Since the ordering of the entropy and chiral condensate determine the
slope, the figure is consistent with the idea that the low temperature
phase contains hadrons, whereas the high temperature phase contains
quarks \cite{quasiquarks}.  At finite $m$ one then has a line of first
order transitions starting at $T=0$ and a large $\mu_B$ and ending
in a critical end point at $(T^E,\mu_B^E)$ which is in the Ising
universality class.  These arguments, based on the Gibbs' phase rule
and the Clapeyron-Clausius equation, reproduce the results of earlier
model studies \cite{kr}, which spurred recent developments in lattice
QCD \cite{lattice} and gave rise to experimental interest in the search
for the QCD critical point \cite{search}.

One can extend such arguments to the other two slopes. Along lines of constant
$\mu_B$ one finds
\beq
   \left.\frac{dT}{dm}\right|_{\mu_B} =
     \frac{|{\mathbf s}_a|-|{\mathbf s}_b|}{S_a-S_b}.
\label{mtslope}\eeq
Given the relative magnitudes of the entropies and chiral condensates, one
finds that this slope is negative, \ie, at constant $\mu_B$, the temperature
of coexistence decreases with increasing quark mass. Along lines of constant
$T$ one finds the slope
\beq
   \left.\frac{d\mu_B}{dm}\right|_T =
     -\frac1B\,\frac{|{\mathbf s}_a|/S_a-|{\mathbf s}_b|/S_b}{1/S_a-1/S_b},
\label{mumslope}\eeq
which is positive. To the best of our knowledge, arguments about these
slopes based on the Clapeyron-Clausius equations are new. They imply that
with increasing quark mass the critical end point, $(T^E,\mu_B^E)$, moves
to larger $T$ and smaller $\mu_B$, as indicated by current lattice data
\cite{sewm}.

\subsection{The section $T$-$\mu_B$-$\mu_I$ when $\Delta m=0$}

The three-dimensional slice through the $N_f=2$ phase diagram which
is obtained for a fixed non-vanishing value of $m$ and $\Delta m=0$
(including the special slice which also has $\mu_B=0$) has been
extensively examined in the literature \cite{ss,kogut,isomu}. This case has
one very attractive feature: the quark determinant is positive definite,
and hence lattice simulations are possible, and they can be used to
check arguments using global symmetries. The drawback is that the symmetry
which makes the quark determinant positive definite is responsible for
making the phase diagram non-generic.

The plane of $T$-$\mu_I$, for arbitrary $m$ and vanishing $\mu_B$
and $\Delta m$,was first discussed in \cite{ss}. Using an effective
theory, it was argued that along the line $T=0$, there should be a phase
transition which results in the development of a charged pion condensate,
${\mathbf p}=\overline\psi\gamma_5\tau_1\psi$.  It was argued that the
phase transition should be critical and in the $O(4)$ universality class
with critical $\mu_I^c=m_\pi$, and that an $O(2)$ critical line emanate
from it.  Now, the Gibbs phase rule does not generically allow a critical
line in a two-dimensional phase diagram. However, subsequent lattice
computations \cite{kogut} saw instead a first order line at finite $T$
connecting to a critical or tricritical point with $\mu_I^c\propto m_\pi$.

The phase diagram was extended to the three dimensional slice
$T$-$\mu_B$-$\mu_I$ for finite $m$ and $\Delta m=0$ by three different
methods in \cite{isomu}. In all these works, critical lines were
obtained in the $\mu_I$-$\mu_B$ surfaces for generic $T$. Thus there
are critical surfaces in this three dimensional phase diagram. Such
a violation of the Gibbs phase rule may occur in subspaces
of high symmetry. For example, in the plane $m=\Delta m=\mu_I=0$, one
found the O(4) critical line shown in Figure \ref{fg.phsdgt0m}. However,
these exceptions are non-generic; when extending the phase diagram they
do not develop into higher dimensional surfaces. This is clear in the
above example. In the same way, one is forced to the conclusion that the
slice $\Delta m=0$ is not generic.  We consider extensions to $\Delta
m\ne0$ next.

\subsection{Isospin broken by quark masses}

\subsubsection{The phase diagram in $T$-$\mu_I$-$\Delta m$}\label{sc.tmuidm}

\begin{figure}[!tbh]
   \begin{center}
   \scalebox{0.3}{\includegraphics{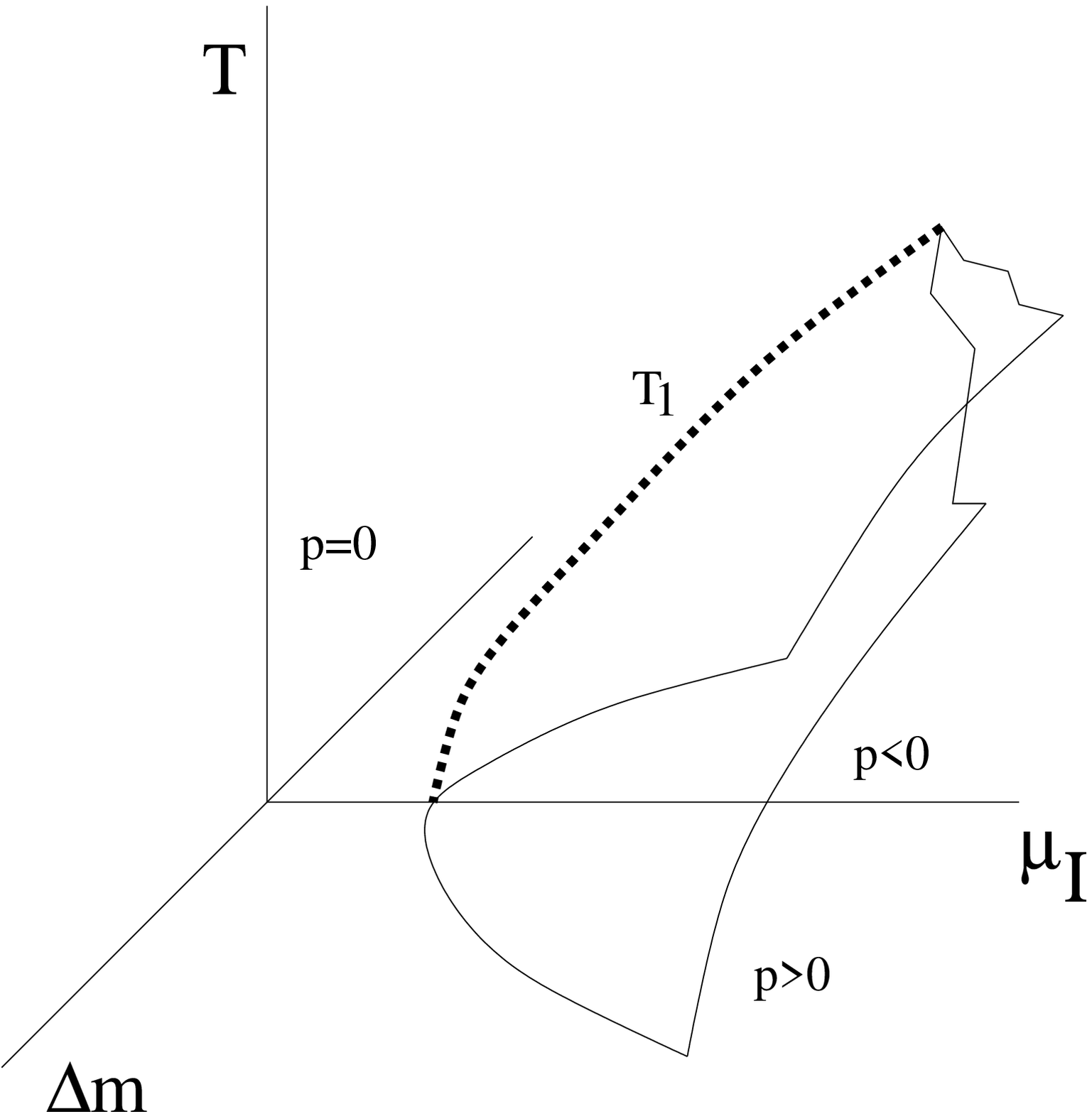}}\hfill
   \scalebox{0.3}{\includegraphics{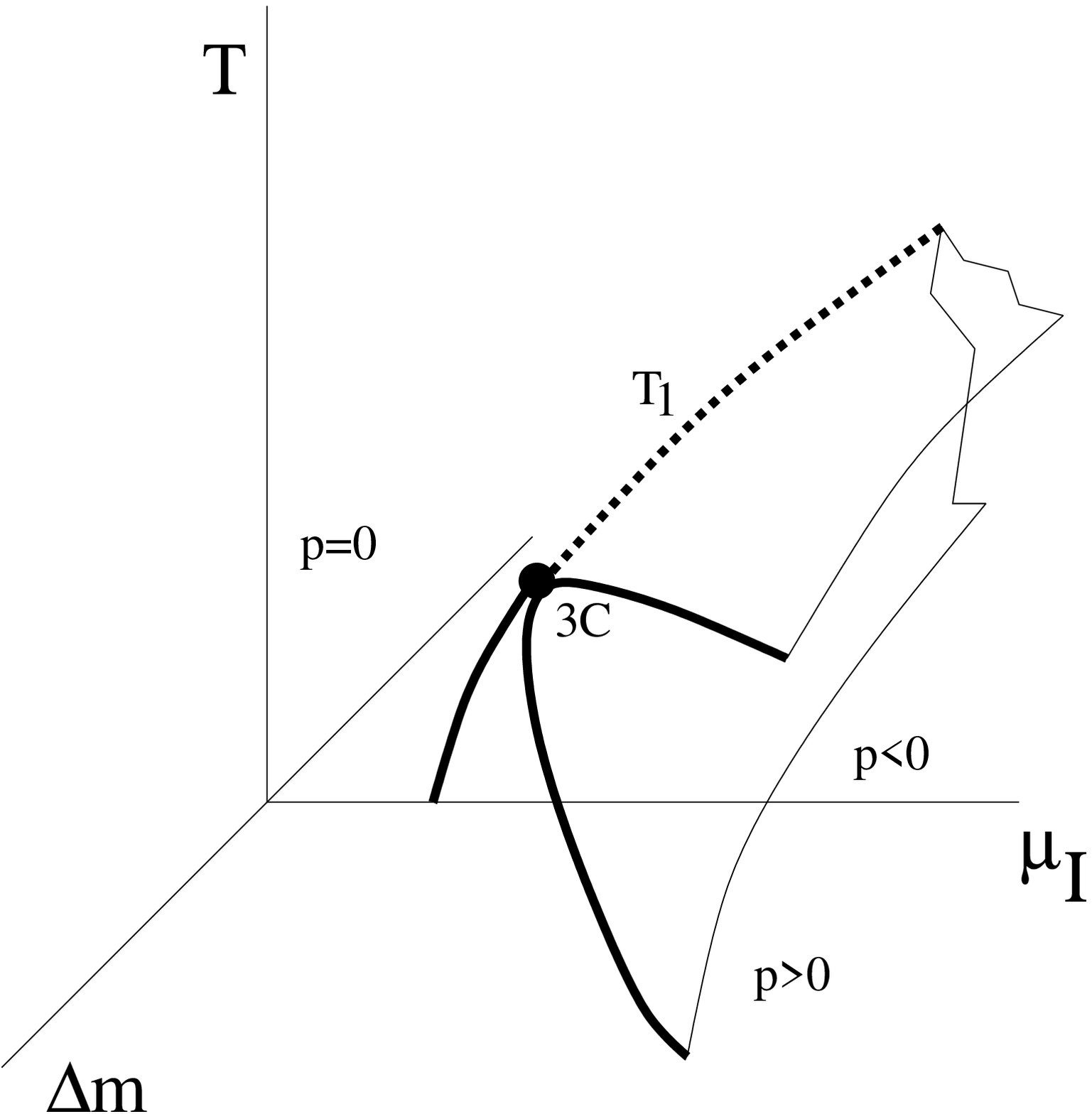}}\hfill
   \scalebox{0.3}{\includegraphics{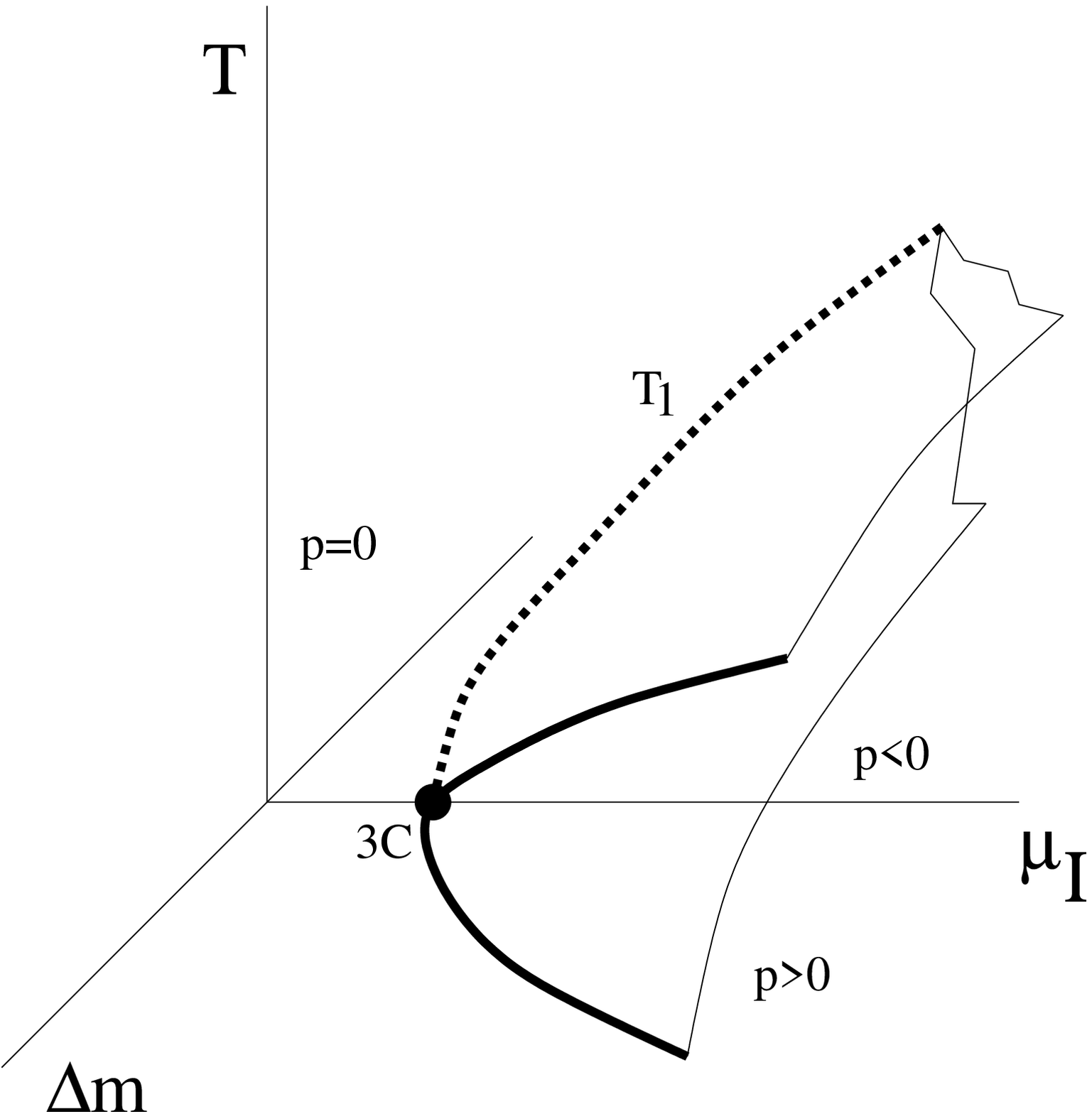}}
   \end{center}
   \caption{The phase diagram in the $T$-$\mu_I$-$\Delta m$ space for
      generic non-zero $m$ and $\mu_B=0$. Since there is a two phase
      coexistence surface on the plane $\Delta m=0$ which separates
      phases with opposite signs of $p$, this plane also contains a
      line of triple-phase coexistence (dashed, marked $T_1$). This
      line may (second and third panels) or may not (first panel) end
      in a tricritical point at $T=0$. The last phase diagram is
      favoured, as we discuss in the text.}
\label{fg.conj2}\end{figure}

Once isospin is broken by a mass difference between the up and down
quarks, an isospin chemical potential no longer matches the up quark and
down antiquark Fermi surfaces exactly. This breaks the symmetry that
kept the full quark determinant non-negative, and hence presents the
usual problems for lattice simulations.  However, it gives a generic
phase diagram in $T$-$\mu_I$.

Consider the section of the phase diagram for generic non-zero $m$ and
$\mu_B=0$.  Changing the sign of $\Delta m$ is equivalent to toggling
the definition of $u$ and $d$ flavours (in the absence of electroweak
interactions) and hence to flipping the sign of $\mu_I$. Thus, if there
is charged pion condensation, \ie, $|{\mathbf p}|>0$, for some value
of $\mu_I$, then there is a first order transition across $\Delta m=0$,
in which the sign of $p$ distinguishes the two phases. Thus, along some
line in the $\Delta m=0$ plane one expects triple-phase coexistence,
the phases corresponding to ${\mathbf p}=0$ and two non-zero values of
${\mathbf p}$ of opposite sign.

As one goes from small to large $\mu_I$ along a line of fixed $\Delta m$
and $T$, the order parameter $\mathbf p$ increases. Hence, by adapting the
argument based on the Clausius-Clapeyron equations for the $T$-$\mu_B$-$m$
phase diagram, one obtains the slope drawn in Figure \ref{fg.conj2}.
Such a 3-phase coexistence line can be accommodated by the three possible
phase diagrams shown in Figure \ref{fg.conj2}.

The first possibility shown in the figure is generic. The triple-phase
line, $T_1$, starts at $T=0$ and continues up. There is no critical
boundary to the three surfaces of first order transitions. This phase
diagram is ruled out by the results of \cite{ss,kogut} which point to
some kind of criticality at $T=\Delta m=0$. However, the $SU_V(2)$
isospin symmetry is broken to a vector U(1) when $\Delta m\ne0$.
As a result, an O(4) critical point is not stable under this perturbation.

The second generic possibility shown in Figure \ref{fg.conj2} takes this fact into
account. The triple-phase line, $T_1$, ends at a tricritical point. Three
critical lines emerge from this point. One lies in the symmetry plane
$\Delta m=0$ and at $T=0$ it is seen to be the O(4) critical
point identified in \cite{ss}. Along the two wing critical lines the
remnant vector $U(1)$ symmetry is broken by charged pion condensation. This
is closely analogous to the phase diagram in Figure \ref{fg.phsdgt0m}. In
this scenario, as one traverses a line of fixed $\Delta m\ne0$ for $T=0$
by varying $\mu_I$, one should go continuously from a phase with small
$\mathbf p$ to one with a large value of this condensate. However, at
$T=0$ and small $\mu_I$ it is not energetically feasible for pions to
condense, hence there must be a range of $\mu_I$ over which $\mathbf p$
vanishes. In that case a phase transition must occur at $T=0$ for any
$\mu_I$.

This brings us to the final possibility: the triple-phase line ends at $T=0$
in a tricritical point. Two wing critical lines emerge from this in
the $T=0$ plane. The third critical line, which is the O(4) line of
the second scenario, is squashed down to a point.  In this scenario
there is always a phase transition between the charged pion condensed
and uncondensed phases. This is the favoured phase diagram, since the
lattice computation of \cite{kogut}, which sees a critical or tricritical
point at $T=0$ and $\Delta m=0$, prefers this over the first scenario.
It would be good to check this through another computation at smaller
lattice spacing.

\subsubsection{The $T$-$\mu_B$-$\Delta m$ phase diagram}\label{sc.tmubdm}

\begin{figure}[!tbh]
   \begin{center}
   \scalebox{0.45}{\includegraphics{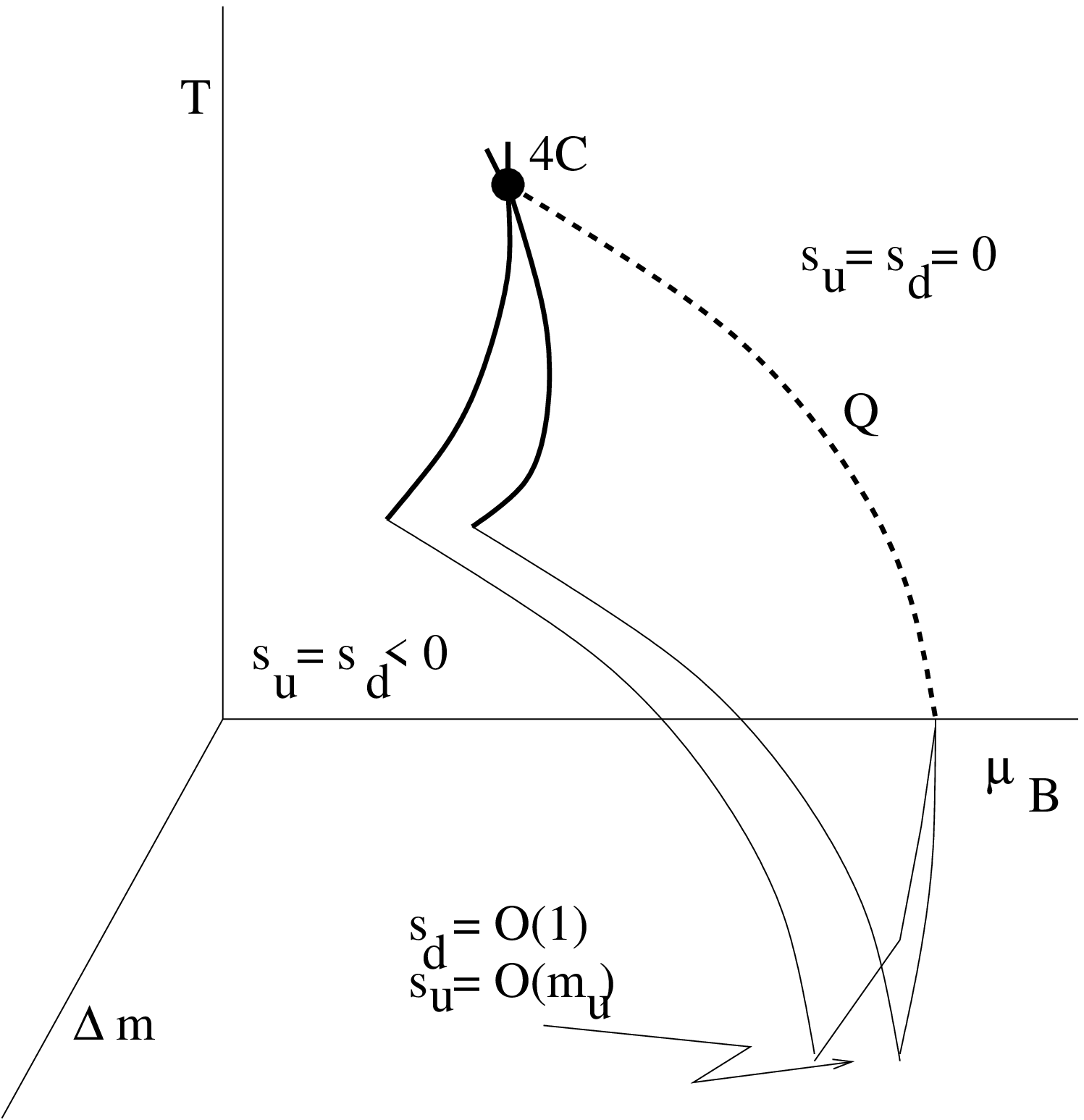}}\hfill
   \scalebox{0.45}{\includegraphics{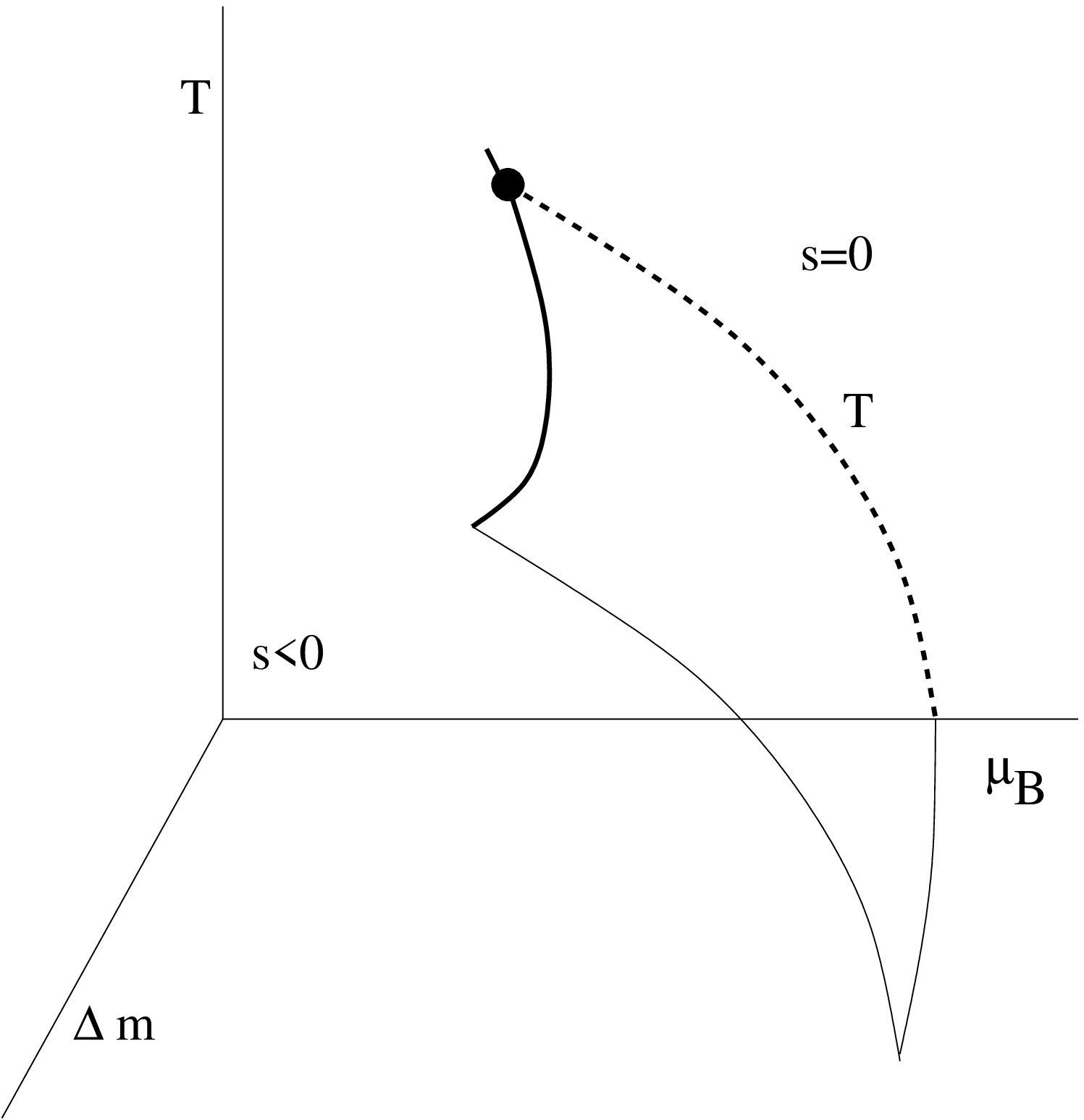}}
   \end{center}
   \caption{Two possible topologies for the phase diagram in
     $T$-$\mu_B$-$\Delta m$ for $\mu_I=0$. The second one is
     preferred (see the text for details).}
\label{fg.imposs}\end{figure}

One might expect that when the isospin symmetry is broken, the flavour
singlet order parameter should be replaced by the pair ${\mathbf s}_u$ and
${\mathbf s}_d$ \cite{isomu}. If so, the $T$-$\mu_B$ phase diagram would
be extended to the first topology shown in Figure \ref{fg.imposs}. This
contains a line of 4-phase coexistence, labeled Q in the diagram with an
end point which is a tetracritical point. In a three dimensional phase
diagram one does not generically expect either a 4-phase coexistence
line or a tetracritical point. We have argued before that non-generic
situations might be obtained on sections of high symmetry. However,
having two different non-vanishing masses breaks the flavour symmetry
maximally, so it is impossible to find a more generic situation.

This no-go argument can be evaded by noting that $\mu_I$ and $\Delta
m$ break the flavour symmetry to the same subgroup. As a result,
it is possible that the topology is the same in the two sections
$T$-$\mu_B$-$\Delta m$ for $\mu_I=0$ and $T$-$\mu_B$-$\mu_I$ for $\Delta
m=0$ separately. In other words, the 4-phase coexistence line and the
tetracritical point lie in the intersection of these two sections in this
extended four dimensional phase diagram $T$-$\mu_B$-$\mu_I$-$\Delta m$.
By the same token, such a topology is ruled out as non-generic in the
$T$-$\mu_B$-$\mu_I$ section for $\Delta m=0$ unless it is also seen in
this section.

Arguing differently, one might expect that for $\Delta m/m<1$ the effect
of isospin symmetry breaking through the quark masses should not change
the phase diagram qualitatively. In that case the two order parameters
${\mathbf s}_u$ and ${\mathbf s}_d$ would be redundant, and the two
sheets of first order transitions merge into a single sheet of first order
transitions found with $\mathbf s$. This possibility is shown as the second
topology in Figure \ref{fg.imposs}.

If this second argument is correct, then one must face the question of
what happens at large $\Delta m/m$. Exactly this question was answered
at $T=0$ using an effective theory in \cite{creutz}.  At finite $\Delta
m$, the explicit breaking of isospin symmetry to vector U(1) allows
the $\pi^0$ to mix with isoscalars (this allows a change in its mass
without changing the masses of $\pi^\pm$).  When $\Delta m/m$ is large,
a neutral pion condensate can form, and, at $T=0$, there is a second
order transition between phases with and without this condensate.
In the real world a mass difference between the charged and neutral
pions is observed, so a mixing of this sort could occur. However, a
neutral-pion condensate is not observed at $T=0$, so one may conclude
that the physical value of $\Delta m/m$ is not large enough for isospin
breaking effects to qualitatively change the phase diagram shown in
the second part of Figure \ref{fg.imposs}.

The mixing of neutral pions with isoscalars is a physical effect
which does not exist in the models used in \cite{isomu}. This leads
us to prefer the second phase diagram shown in Figure \ref{fg.imposs}.
The only lattice study to date of thermodynamics at finite $\Delta m$
\cite{flavbrk} saw no evidence of two separate crossovers. A computation
with a different lattice formulation of quarks will be an interesting
and useful additional piece of evidence.

\subsection{The physical phase diagram}\label{sc.physical}

\begin{figure}[!tbh]
   \begin{center}
   \scalebox{0.7}{\includegraphics{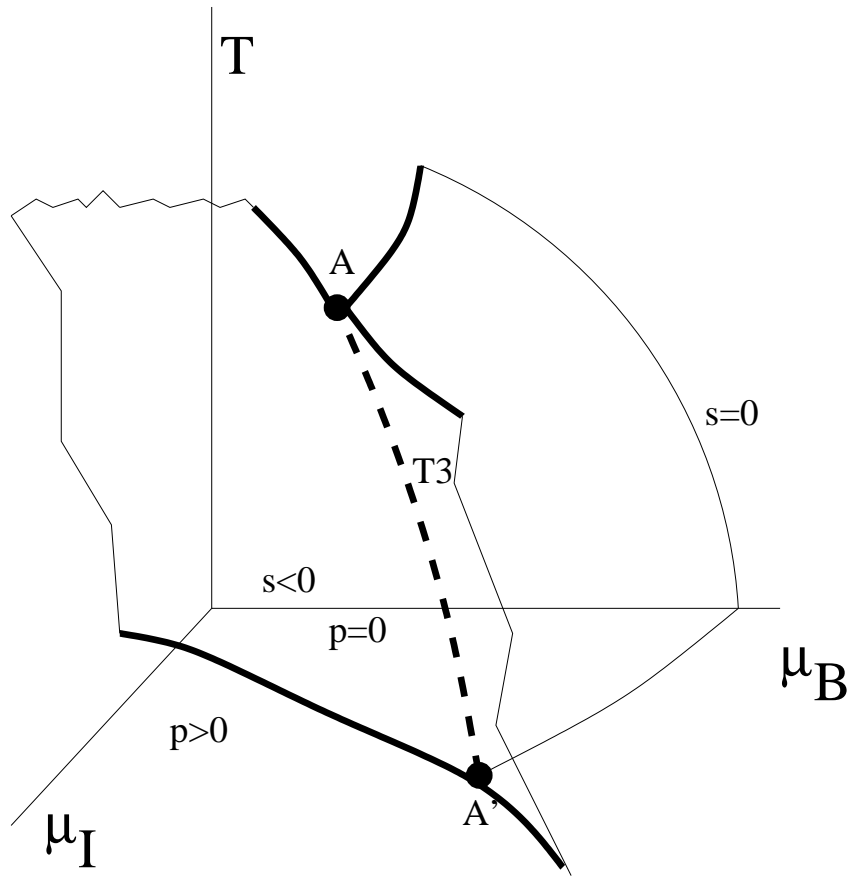}}
   \end{center}
   \caption{The phase diagram in the $T$-$\mu_B$-$\mu_I$ space
      for generic non-zero $m$ and $\Delta m$ is organized by the
      two order parameters $\mathbf s$ and $\mathbf p$. It contains
      a line of 3-phase coexistence (labeled $T3$) and two tricritical
      points (labeled $A$ and $A'$).}
\label{fg.conj3}\end{figure}

The physical phase diagram of QCD is the one in which all possible
experimentally tunable parameters are shown for fixed physical values
of the remaining parameters. This is the phase diagram in the space of
$T$-$\mu_B$-$\mu_I$ for generic small $m$ and $\Delta m<m$. After the
analysis of the previous subsections, deducing the topology of this phase
diagram is straightforward. The various phases are distinguished by the
order parameters $\mathbf s$ and $\mathbf p$.  We discuss the topology of
the phase diagram (shown in Figure \ref{fg.conj3}) only in the quadrant
with $\mu_{B,I}>0$. The remainder can be constructed by symmetry.

In the $\mu_I=0$ plane there is a first order line ending in the QCD
critical point. Across this line $\mathbf s$ changes discontinuously.
Following the discussion of the previous subsection, we conclude that
this line develops into a surface of 2-phase coexistence. The QCD critical
end point develops into a critical line (in the liquid-gas universality class)
which is the boundary of this surface.  In the $\mu_B=0$ plane there is
a first order line separating phases with vanishing and finite values
of $\mathbf p$ with a liquid-gas critical point at $T=0$. This develops into
a 2-phase coexistence surface.

The two first order surfaces must meet along some line, $T3$. Hence
$T3$ is a line of 3-phase coexistence, and another surface of 2-phase
coexistence must emanate from it, separating the phase in which both
the condensates vanish from that in which neither does. The two ends
of $T3$, labeled $A$ and $A'$, must both be tricritical points. At $A$
the first order surface of discontinuity in $\mathbf p$ must have a
boundary. Since the system has no particular symmetry at $A$, all three
critical lines meeting at this point must be in the universality class of
a liquid-gas transition.  At $A'$ there seem to be two critical lines,
both in the liquid-gas universality class. The third critical line is
degenerate, as in the third topology shown in Figure \ref{fg.conj2}.

The derivation of the Clapeyron-Clausius equations for the physical
phase diagram needs little further comment. Along planes of constant $\mu_B$
one can use eq.\ (\ref{mutslope}), since $m$ and $\Delta m$ are
also constant along this line. Along the planes where $\mu_I$ are
constant, the modification of eq.\ (\ref{mutslope}) was already
discussed in Section \ref{sc.tmuidm}, and leads to the opposite slope
of the coexistence surface for pion condensation. The remaining part
involves scaling the expression $d{\cal G}_i = S_i dT+B d\mu_B+{I_3}d\mu_I$
by $S_i$ in the phase $i$, and equating the two free energy densities so
defined at each coexistence surface. The resulting Clapeyron-Clausius equation
is
\beq
   \left.\frac{d\mu_B}{d\mu_I}\right|_T = -\frac{I_3}B.
\label{mubmutcceq}\eeq
The first order surface of chiral symmetry restoration, which is in the
phase without a pion condensate, then has only weak dependence on $\mu_I$.
Similarly, the first order surface of pion condensation has weak dependence
on $\mu_B$ because the baryon density is low in the phase with chiral
symmetry breaking. The surface separating the phase with both condensates
vanishing from the phase where both condensates are present has more
complicated behaviour. One expects that in the phase without condensates
the number densities scale with chemical potentials as $B\propto\mu_B^3$
and $I_3\propto\mu_I^3$ when $T$ is small. In that case this first order
surface is roughly linear in $\mu_B$ and $\mu_I$. Lattice computations
at $\Delta m=0$ can substantially improve our knowledge of the functional
form of $B(T,\mu_B,\mu_I)$ and $I_3(T,\mu_B,\mu_I)$, and therefore the
shape of the physical phase diagram through continuation to $\Delta m\ne0$.

The position of $A$ and $A'$ are constrained by these considerations.
It is estimated that the QCD critical end point at $\Delta m=\mu_I=0$ is
at $\mu_B^E/T^E\simeq1$ with $T^E\simeq170$--$190$ MeV. Since the effect
of isospin breaking is small, one expects that at realistic values of
$\Delta m$ the critical end point does not shift drastically. One also
knows that at $\mu_B=\Delta m=0$, the critical point in $\mu_I\simeq
m_\pi$. The corrections at realistic $\Delta m$ are not large.
One expects that $\mu_B^A\simeq\mu_I^A\simeq T^A\simeq100$--200
MeV. Also, from this topology one could deduce that in heavy-ion collisions
where $\mu_I\ll m_\pi$ one should expect to see the QCD critical line close
to the estimates of $\mu_B^E$ and $T^E$.

\section{Three flavour}

\begin{figure}[!tbh]
   \begin{center}
   \scalebox{0.5}{\includegraphics{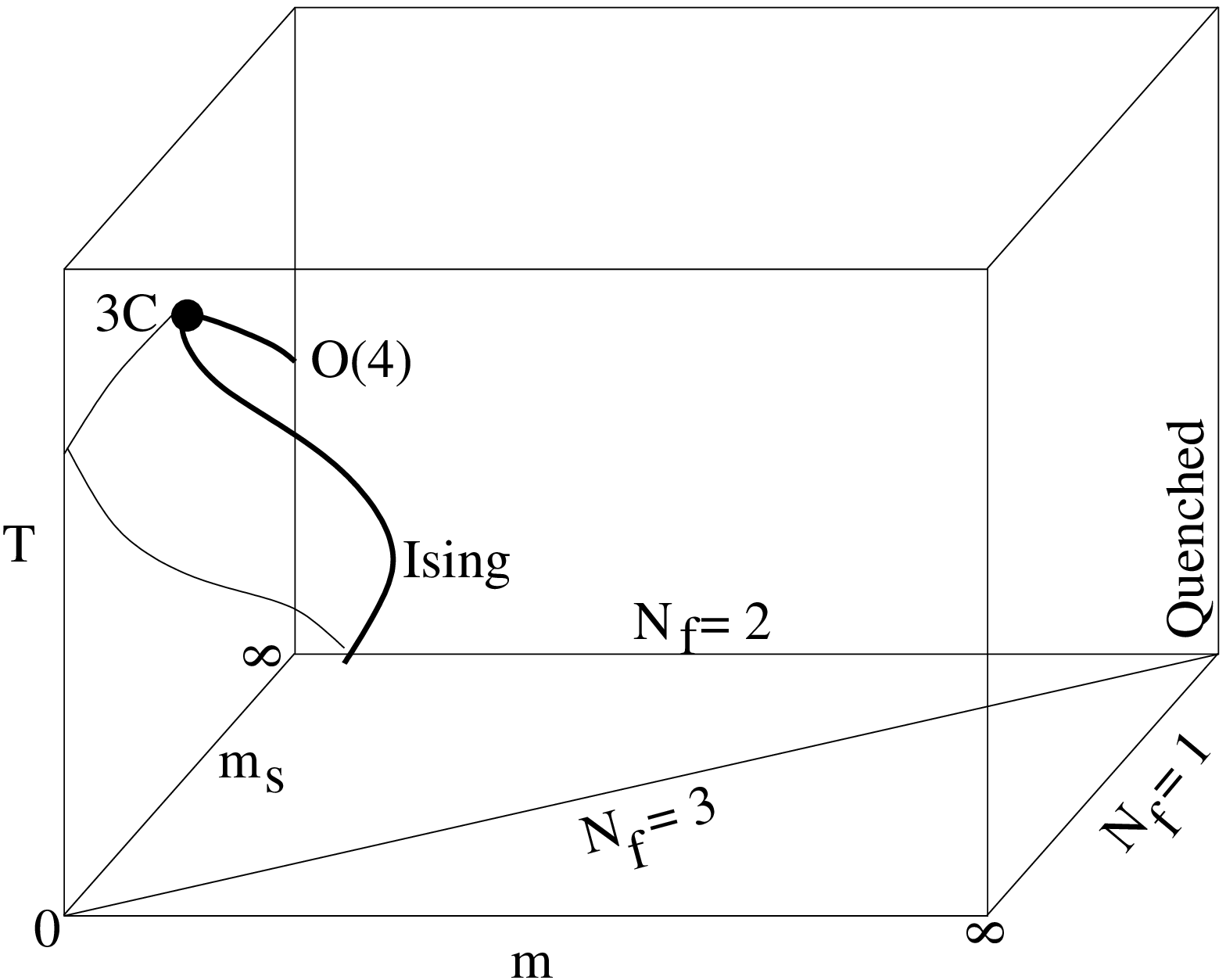}}\hfill
   \scalebox{0.6}{\includegraphics{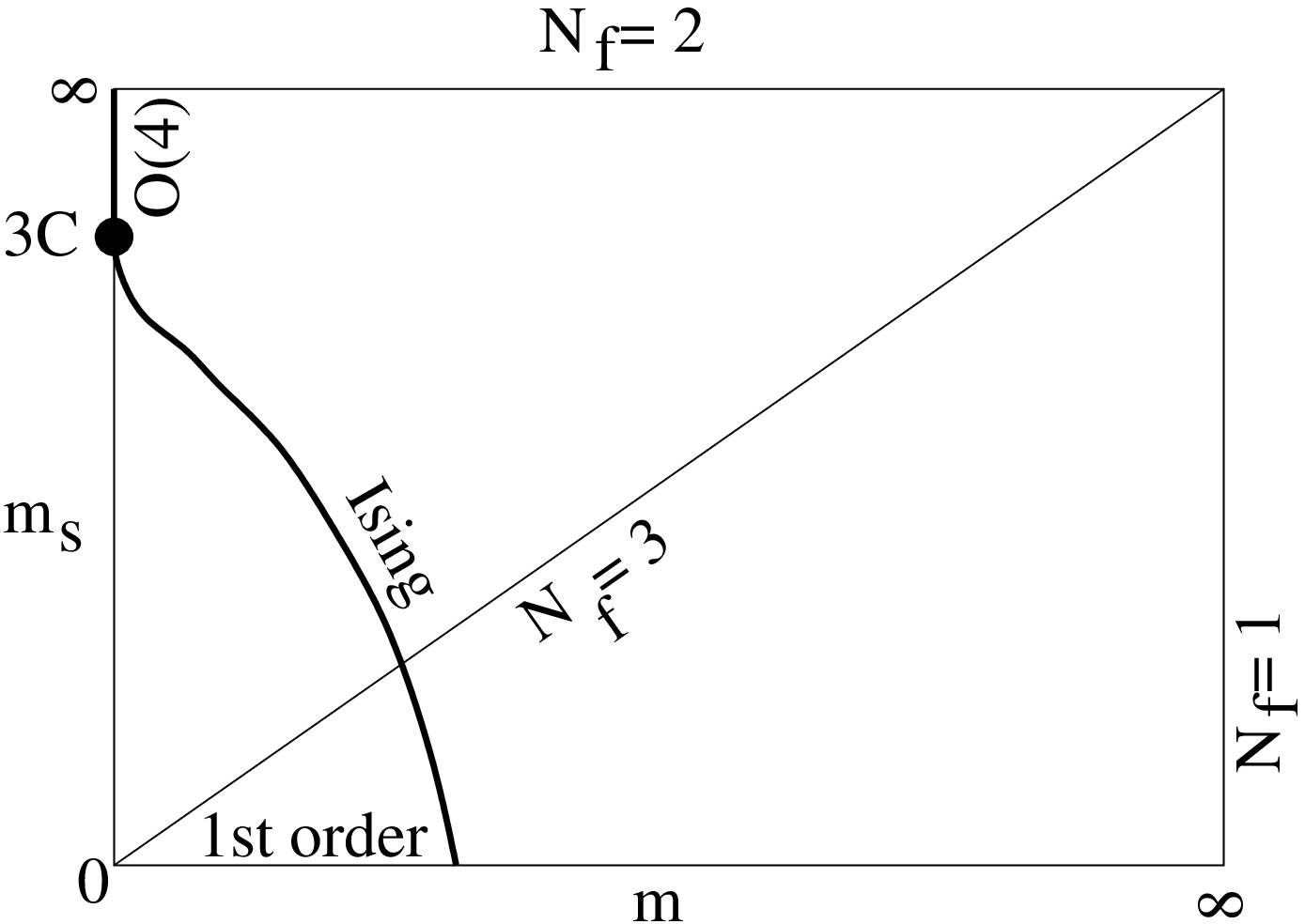}}
   \end{center}
   \caption{The phase diagram of QCD for vanishing chemical potentials and
      $\Delta m=0$ (on the left) as observed with the light quark condensate.
      In different parts of this phase diagram
      different flavour symmetries are obtained: the diagonal plane has
      exact 3-flavour symmetry, the ``back'' plane has $N_f=2$, the far
      corner is the quenched theory, and the plane at the far right has
      $N_f=1$. On the right is the corresponding flag
      diagram, obtained by projecting the phase diagram to the plane of
      $T=0$.}
\label{fg.flag}\end{figure}

For $N_f=3$ QCD, the intensive quantities at the phase transition are
the temperature $T$, the three quark masses, $m_u$, $m_d$ and $m_s$, and
the three chemical potentials $\mu_u$, $\mu_d$ and $\mu_s$. As before,
we can take linear combinations of these variables to parametrize
the seven-dimensional phase diagram. We will use the mean light quark
mass $m=(m_u+m_d)/2$ and the mass splitting between these two, $\Delta
m=m_u-m_d$. It is useful to construct the baryon chemical potential,
$\mu_B$ and two other combinations, as outlined in \cite{quasiquarks}.

The three flavour phase diagram has barely been explored, and we shall
not be able to do justice to the many kinds of pairings that it may
possess. In this first work we restrict attention to a question which
has been asked recently: is the two-flavour phase diagram for chiral
symmetry restoration (Figure \ref{fg.phsdgt0m}) a good guide to the
corresponding phase diagram in QCD with additional strange quarks?

\subsection{Vanishing chemical potentials}

Universality arguments \cite{pw} lead us to expect that the chiral phase
transition in three flavour QCD at zero chemical potentials is of first
order. The order parameter used is the three-flavour condensate. In the
three-flavour chiral limit, since the action is blind to flavour, any
linear combination of three flavoured condensates is as good an order
parameter as any other, so one could as well continue to use the two-flavour
order parameter $\mathbf s$ to study the phase structure. When the strange
quark becomes massive with the light quarks remaining massless, one does
not expect symmetry restoration in the strange sector while the symmetry
remains broken in the light quark sector. Similar expectations hold whenever
the strange quark mass is heavier than the up and down quark masses.  Hence,
in this whole region one expects that the same phase structure is seen whether
one uses the three flavour condensate or $\mathbf s$.

The phase diagram of QCD with three quarks has been explored in several
lattice computations \cite{3flav,columbia,ising,wuppertal} where the
two light quarks are degenerate, \ie, $\Delta m=0$. In this case the
phase diagram in the section $m$-$m_s$-$T$ is of the kind presented in
Figure \ref{fg.flag}. In these studies there is only one chiral phase
transition observed, and not separate ones for strange and light flavours,
in conformity with the above discussion. For infinite $m_s$ and $m=0$,
one has two flavour chiral symmetry, and hence a second order transition
in the O(4) universality class. When $m_s=m$, \ie, three flavour symmetry
is exact, one has a line of first order transitions up to a critical end
point, which lies in the Ising universality class \cite{ising}. Moving
off the exact $N_f=3$ line (termed going to $N_f=2+1$ in the literature),
this develops into a surface of first order transitions bounded by a
critical line. Since this line is in the liquid-gas universality class
\cite{ising}, it must meet the O(4) line in a tricritical point. The
nature of this critical line is crucial in establishing a phase
diagram. If it is in the liquid-gas class, as expected, then this would
be visible with any formulation of lattice quarks.

\begin{figure}[!tbh]
   \begin{center}
   \scalebox{0.5}{\includegraphics{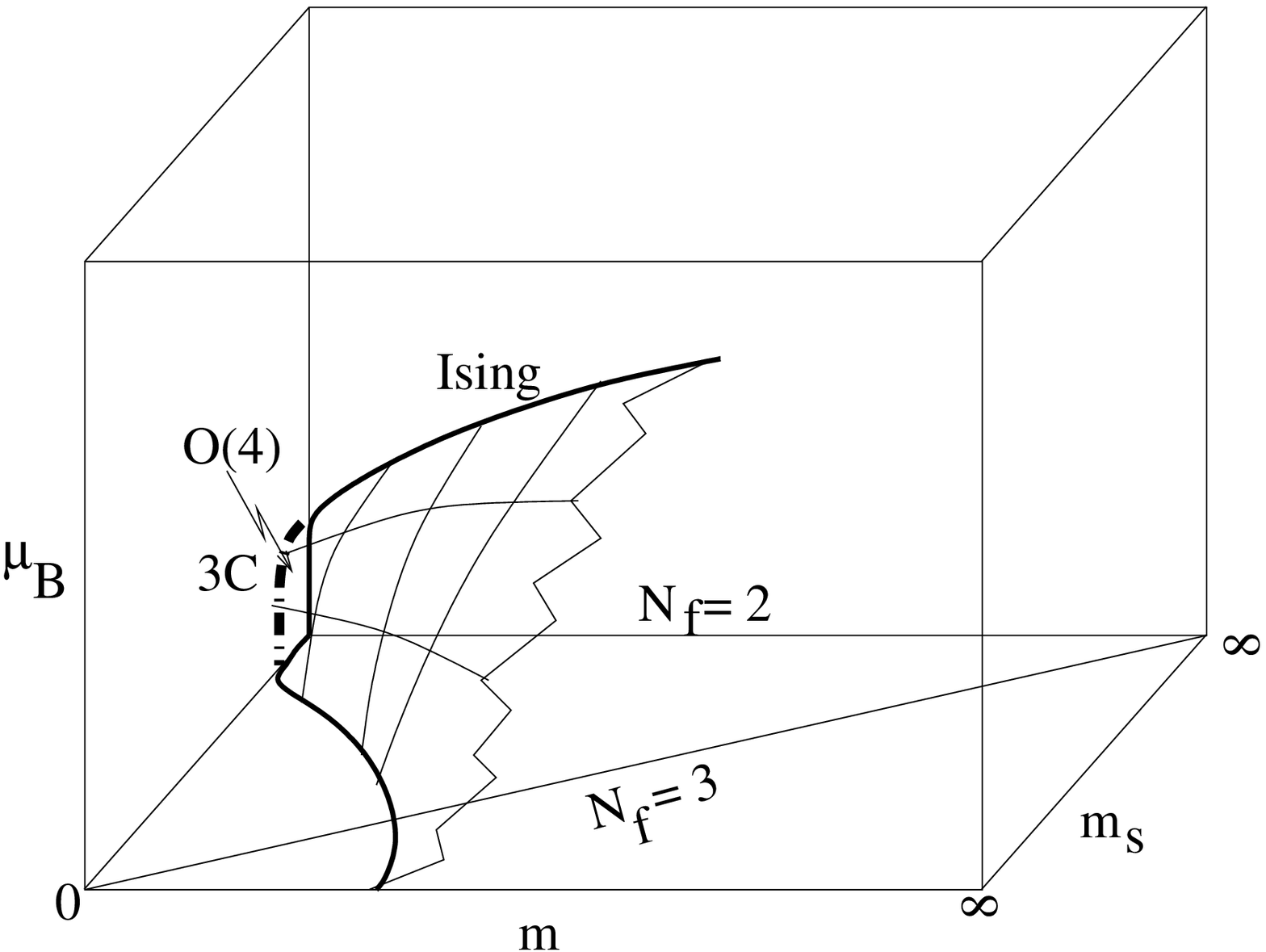}}\hfill
   \scalebox{0.5}{\includegraphics{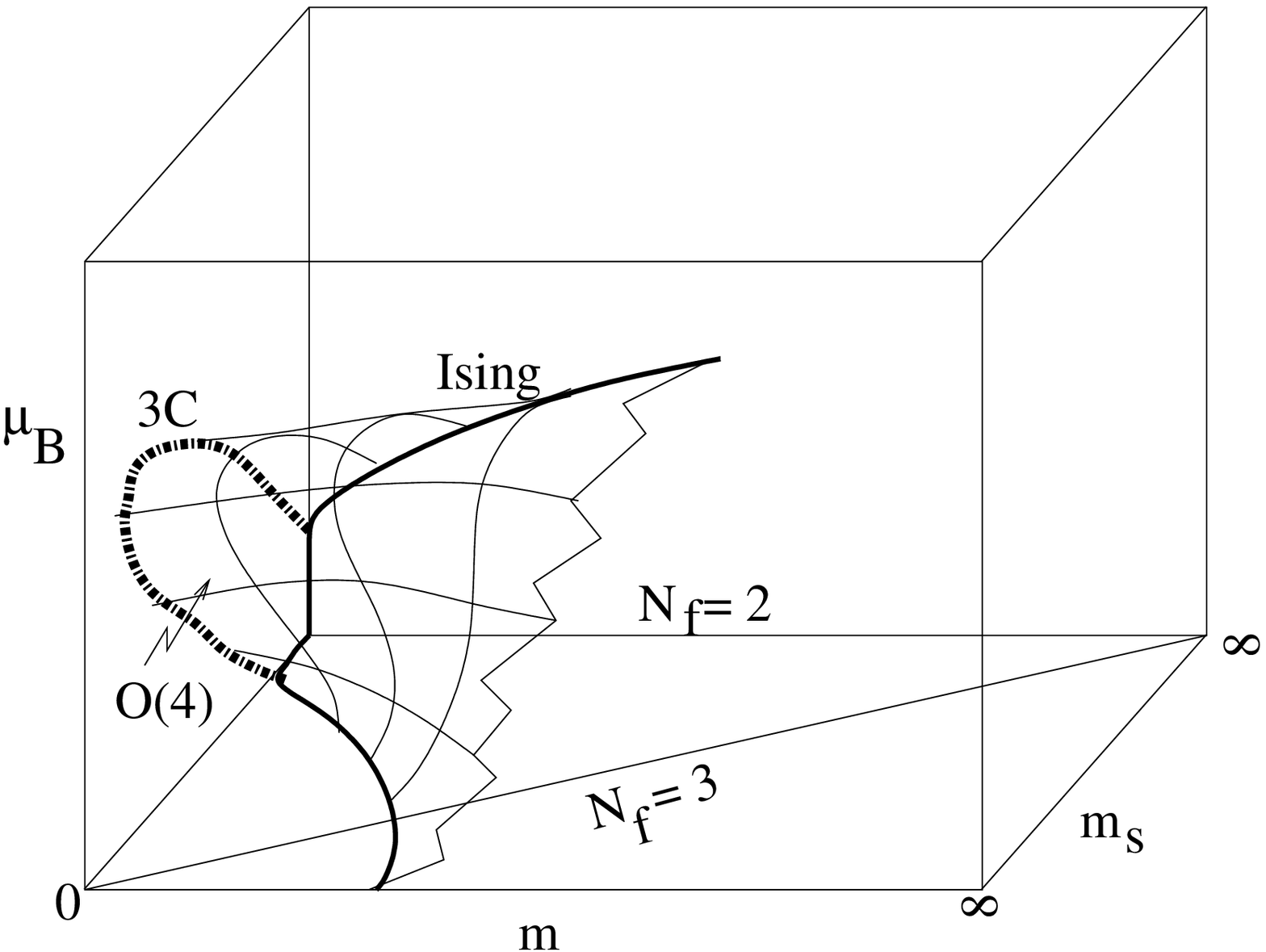}}
   \end{center}
   \caption{Possible flag diagrams for three flavour QCD are organized
      around the shape of the tricritical line (dash-dotted line). The
      surface of O(4) criticality which lies in the plane with $m=0$
      meets a surface of Ising criticality along this line. The
      two possibilities shown are distinguished by the approach of the
      tricritical line to the $\mu_B=0$ plane. The second possibility
      includes the case where the tricritical line goes to infinite
      $\mu_B$ somewhere between its two fixed ends.}
\label{fg.possible}\end{figure}

The phase diagram shown in Figure \ref{fg.flag} can be probed entirely with
the light quark condensate. There are, of course, other
phase transitions in this phase diagram--- for example the deconfining
transition in the corner which contains the quenched theory. This is
observed with a different order parameter, the Polyakov loop, and hence
does not appear in the phase diagram shown.

These results are often presented in a figure (the second of Figure
\ref{fg.flag}) which is the phase diagram projected down to the plane
$m$-$m_s$. This is not a phase diagram since a point on it does not
represent a stable thermodynamic phase.  Instead, each point on this
diagram represents whether or not there is a phase transition at some
$T$ ``above'' it. Unfortunately there seems to be no name for such an
useful diagram, and we are forced to invent a name for it. In the rest
of this paper we call it the ``flag diagram'', because each point flags
whether there is a phase transition above it, and if so, the order of the
transition.

One of the main open questions about the flag diagram is the location
of the physical point. All computations indicate that this lies deep
in the crossover region \cite{wuppertal,biswa}.  A second important
question is about the location of the tricritical point. What is
the value of $m_s^{3c}$?  If the physical strange quark mass is
less than $m_s^{3c}$ then the $N_f=2$ phase diagram will be a good
guide to the phase diagram of real QCD. If, on the other hand, the
physical strange quark mass is larger, then the phase diagram in the
real world could be significantly different.  One naively expects that
$m_s^{3c}\simeq\Lambda_{{\scriptscriptstyle QCD}}$; since the physical
value of $m_s$ is also similar, this is an interesting and wholly
non-perturbative question.

One estimate of $m_s^{3c}$ that we are aware of was made
using a linear sigma model with parameters fixed by hadronic data
\cite{szep}. This finds that the K meson mass at the tricritical point
is at least 1700 MeV, far in excess of the physical value. While this
would seem to indicate that the physical strange quark mass is lighter
than $m_s^{3c}$, there are indications that further work may be needed
in order to pin down $m_s^{3c}$. For example, in \cite{szep} it is
estimated that the critical end point along the $N_f=3$ line occurs at
$m_\pi^c=m_K^c=110\pm20$ MeV. The lattice computations of \cite{ising}
indicate that the critical end point may be at $m_\pi^c=m_K^c=192\pm25$
MeV. Whether the discrepancy is due to an insufficiency in the model of
\cite{szep} or the multiple extrapolations performed in \cite{ising},
or a combination of the two, remains to be tested in future. In \cite{pf}
the estimate $m_s^{3c}\simeq500$ MeV is presented. Indications that the
cutoff effects at comparable cutoffs are large come from the results of
\cite{wuppertal,biswa}.

\begin{figure}[!tbh]
   \begin{center}
   \scalebox{0.7}{\includegraphics{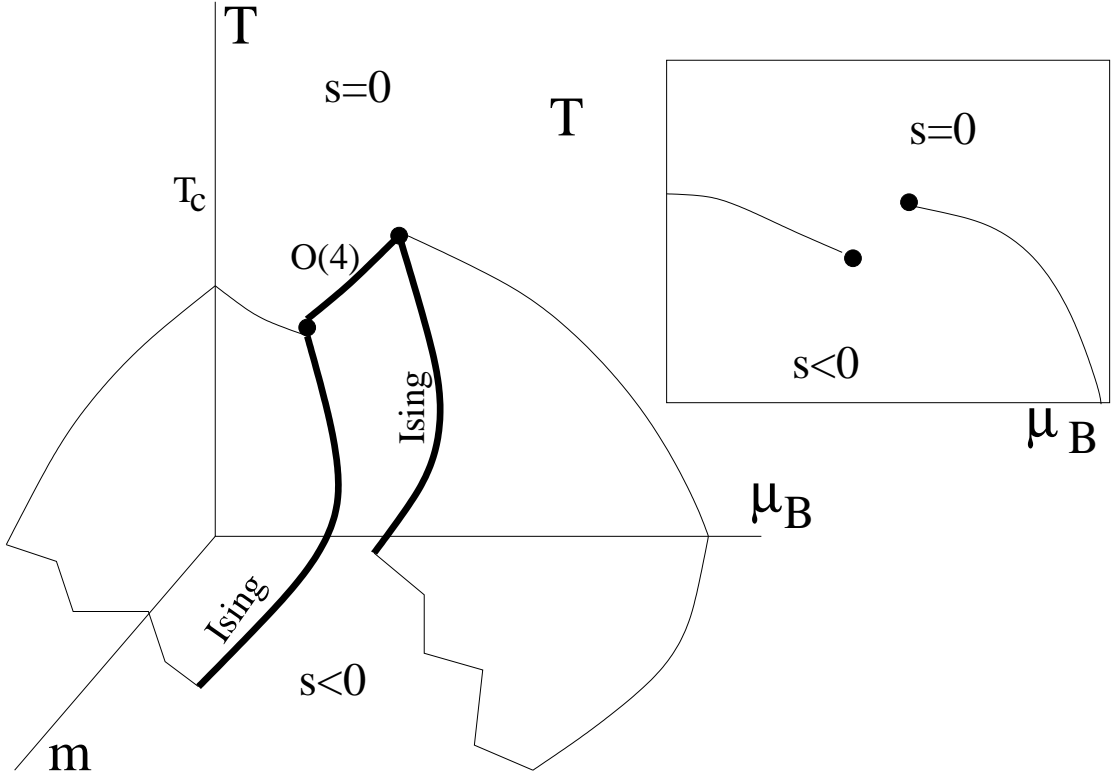}}
   \end{center}
   \caption{The phase diagram for fixed $m_s^{turn}<m_s<m_s^{3c}$ and
      $\Delta m=\mu_I=\mu_Y=0$ in the second scenario shown in Figure
      \ref{fg.possible}. The phase diagram for fixed $m$ is shown in
      the inset.}
\label{fg.unusual}\end{figure}

\subsection{Finite $\mu_B$}\label{sc.phasd}

Another slice that is interesting from the point of view of forthcoming
planned experiments is the four-dimensional $T$-$\mu_B$-$m$-$m_s$ slice
obtained with $\Delta m=\mu_I=\mu_Y=0$. This was investigated recently
on the lattice \cite{biswa,pf}, and a flag diagram obtained by projecting
out the $T$ axis was used in the presentation. These two studies differ
in the conclusions reached: one \cite{biswa} favours the first flag
diagram shown in Figure \ref{fg.possible}; the other \cite{pf}, we
argue below, is consistent with the second possibility shown in the same
figure. These are computations which present many technical challenges. We
remark later on one way to resolve the issue.

Two sections of this phase diagram have been discussed--- the section
with $\mu_B=0$ was discussed in the previous subsection, and the section
with $m_s=0$ was discussed in the section on the $N_f=2$ phase diagram
in $T$-$\mu_B$-$m$. Both contain a line of O(4) critical points, and
a line of Ising critical points, meeting at a tricritical point. The
two lines of O(4) critical points are in fact the same, since they both
originate in the finite-$T$ transition for the $N_f=2$ chiral symmetry
restoration. In the 4-dimensional phase diagram one should, therefore,
see a single critical surface of O(4) transitions. The boundary of this
critical surface is the tricritical line, the two ends of which show
up as the two tricritical points we have already discussed. At this
tricritical line the O(4) critical surface is glued on to surfaces
of Ising criticality.  The tricritical line can also be viewed as the
boundary of a surface of triple phase coexistence. This much is a direct
consequence of the Gibbs phase rule.

The Gibbs phase rule also allows a tetracritical point in a generic four
dimensional phase diagram. A tetracritical point is the end point of a
line of four-phase coexistence. With a single scalar order parameter,
such as $\mathbf s$, it is not possible to distinguish four phases.
Therefore, we conclude that a tetracritical point should not be seen in
the four dimensional phase diagram observed with $\mathbf s$.

There remain two generic possibilities for the shape of the phase diagram
and they are shown in terms of the corresponding flag diagram in Figure
\ref{fg.possible}. The first possibility is that the tricritical line
lies entirely on the side $m_s\ge m_s^{3c}$. The second is that the
tricritical line continues to some $m_s^{turn}<m_s^{3c}$ and then bends
back towards $m_s^{3c}$. In the first scenario, the critical line moves
to larger $m$ and $m_s$ as $\mu_B$ is increased by a small amount from
$\mu_B=0$.  In the second, the opposite movement occurs until a large
value of $\mu_B$.  This is precisely the observation which distinguishes
the two lattice studies in \cite{biswa,pf}. It would be interesting to
push lattice studies to smaller cutoffs and decide which scenario is
actually obtained in QCD.

In the first case the phase diagram for $m_s>m_s^{3c}$ is qualitatively
like the $N_f=2$ phase diagram, and for $m_s<m_s^{3c}$, similar to
$N_f=3$. In the second case for $m_s^{3c}>m_s>m_s^{turn}$ there are two
sheets to the phase diagram at fixed $m_s$: one similar to the $N_f=2$
phase diagram, the other like the $N_f=3$ phase diagram, with the two
tricritical points joined by an O(4) critical line, as shown in Figure
\ref{fg.unusual}. The inset in the figure shows the unusual phase diagram
for fixed non-zero value of $m$. 
It would be very interesting if QCD is eventually
seen to yield such a phase diagram.

\section{Conclusions}

The phase diagram of QCD is complicated. The large global symmetries
of QCD imply that there are many thermodynamical intensive parameters,
\ie, free parameters which enter the Lagrangian. As a result, there
are many different kinds of pairings of quarks that can arise as these
parameters are changed. In this paper we have considered only small
chemical potentials, defined to be the region where the relevant pairings
are between a quark and an antiquark.

The topology of the phase diagram for QCD with two flavours of quarks
has been investigated earlier by many authors, mainly in sections of
partial symmetry.  In this paper we have extended these considerations
to the case of physical quark masses, which break both the chiral
symmetry and isospin symmetry.  Explicit breaking of isospin symmetry
by quark masses leads to mixing of isovectors with isoscalars and hence
influences the phase diagram (see Section \ref{sc.tmubdm}).  This piece
of physics is incorporated into the phase diagram that we construct.
The topology of this physical phase diagram is strongly constrained
by thermodynamics--- essentially the Gibbs phase rule (see Section
\ref{sc.gibbs}). Our results are given in Section \ref{sc.physical} and
summarized in the phase diagram shown in (see Figure \ref{fg.conj3}). The
physical phase diagram contains a triple line at which phases with and
without the chiral and pion condensates coexist.  Deducing actual values
of the critical and tricritical points in this diagram is a much harder
task. However, appropriate Clapeyron-Clausius equations along with the
results of current lattice computations \cite{lattice} place various
constraints on them.  Improving these constraints through future lattice
computations is possible without extreme effort.

The phase diagram of three flavour QCD could be more complicated, since
there are even larger global symmetries. Strange quarks can decay into
the light quarks through the weak interactions, so one may ask whether
this phase diagram is relevant to any physical system. In the context
of heavy-ion collisions, where the system is expanding so fast that
weak interactions do not come into play in the thermodynamics, strange
quarks are indeed relevant. Further, in the physics of compact stars
(involving large chemical potentials), where weak decays of strange
quarks are suppressed due to Fermi-blocking of the light quark states
into which they could have decayed, strange quarks also become relevant
to thermodynamics. Thus the study of the three flavour phase diagram
is interesting.

Comparatively little is known at present about this phase diagram. Here we
restrict attention to the breaking and restoration of chiral symmetry. The
Gibbs phase rule places strong constraints on the topology of the phase
diagram.  As the strange quark mass is increased keeping the light quarks
massless, there is a tricritical value, $m_s^{3c}$, at which the finite
temperature chiral phase transition changes from being first order to
second order (for $m_s>m_s^{3c}$). Presently available observations on
the lattice \cite{biswa,pf} can be patched together into two possible
phase diagrams which we have discussed in Section \ref{sc.phasd}. The
three-flavour phase diagram can be improved substantially with lattice
computations in the near future.

Apart from the results on the QCD phase diagram we believe that this
paper demonstrates an useful technique. Usually phase diagrams have been
examined by writing down a Ginzburg-Landau theory. As remarked before,
these give useful predictions for universal quantities. However, when
all symmetries are broken, the usefulness of such a theory is severely
curtailed. In such situations, we found that the Gibbs' phase rule is
a handy tool for exploring the topology of phase diagrams. When
supplemented with the Clapeyron-Clausius equations, it may even be
possible to make some quantitative statements of the kind that have
been explored in this work.

It is a pleasure to thank Mike Creutz, Saumen Datta, Philippe de Forcrand,
Rajiv Gavai, Bengt Petersson, and Misha Stephanov for discussions. This
work was supported by the Indian Lattice Gauge Theory Initiative.

\end{document}